\documentclass[aps,preprint,superscriptaddress,superscriptemail,showpacs,showkeys]{revtex4}
\usepackage[dvipdf]{graphicx}
\usepackage{amssymb}
\usepackage{amsbsy}
\usepackage{latexsym}

\newcommand{\be}{\begin{equation}}
\newcommand{\ee}{\end{equation}}
\newcommand{\ba}{\begin{eqnarray}}
\newcommand{\ea}{\end{eqnarray}}

\begin{document}

\title{Massive logarithmic graviton \\ in the critical generalized  massive gravity}

\author{Yong-Wan Kim}

\email{ywkim65@gmail.com}

\affiliation {Center for Quantum Spacetime, Sogang University,
Seoul 121-742, Korea}

\author{Yun Soo Myung}

\email{ysmyung@inje.ac.kr}

\affiliation {Institute of Basic Science and School of Computer
Aided Science, Inje University, Gimhae 621-749, Korea}

\author{Young-Jai Park}

\email{yjpark@sogang.ac.kr}

\affiliation {Center for Quantum Spacetime, Sogang University,
Seoul 121-742, Korea}

\affiliation {Department of Physics and Department of Global Service Management,\\
Sogang University, Seoul 121-742, Korea}

\begin{abstract}

We study the generalized  massive gravity in three dimensional
flat spacetime. A massive logarithmic mode is propagating in the
flat spacetime at the critical point where two masses degenerate.
Furthermore,  we discuss the logarithmic extension of the Galilean
conformal algebra (GCA) which may arise from the exotic and
standard rank-2 logarithmic conformal field theory (LCFT) on the
boundary of AdS$_3$ spacetime.
\end{abstract}



\maketitle

\section{Introduction}

Recently, a holographic correspondence between a conformal
Chern-Simon gravity (CSG) in the flat spacetime and a chiral
conformal field theory  was reported in~\cite{Bagchi:2012yk}.
Choosing  the CSG as the flat-spacetime limit of the topologically
massive gravity  in the scaling limits of $\mu\to 0$ and $G\to
\infty$ while keeping fixed $\mu G$, two Bondi-Metzner-Sachs (BMS)
central charges $c_1$ and $c_2$ are determined to be $c_1=24k $
and $c_2=0$.  The CSG is conjectured to be dual to a chiral half
of a CFT with $c=24k$. This establishes the BMS/GCA
correspondence~\cite{Bagchi:2010eg,Barnich:2006av,Bagchi:2009pe}
when choosing a concrete model of the CSG.

The linearized equation of the CSG leads to the third order equation
$D^3h=0$ in the Minkowski spacetime.  The  solution to the
first-order massless equation $Dh^\xi=0$ is given by $
h^\xi_{\mu\nu}=e^{-i(\xi+2)\theta} r^{-(\xi+2)}(m_1\bigotimes
m_1)_{\mu\nu}$ in Ref.~\cite{Bagchi:2012yk}, where $\xi$ is the
eigenvalue of GCA generator  $L_0$ and $m_1$ is an ISO (2,1)
generator
$m_1=ie^{i\theta}(\partial_u-\partial_r-i\partial_\theta/r)$.
Furthermore, the solution to $D^2h^{\rm log}=0$ is given by $ h^{\rm
log}_{\mu\nu}=-i(u+r) h^\xi_{\mu\nu}$, while the solution to
$D^3h^{\rm log^2}=0$ is $h^{\rm log^2}_{\mu\nu}=-\frac{1}{2}(u+r)^2
h^\xi_{\mu\nu}$. Even though $\{h^{\rm log},h^{\rm log^2}\}$ do not
include logarithmic terms, they were considered as the
flat-spacetime analogues of log- and log$^2$-solutions on the
AdS$_3$ spacetime. We would like to mention that the solutions
$\{h^\xi,h^{\rm log},h^{\rm log^2}\}$ could not represent any
physical modes propagating on the flat spacetime background because
the CSG  has no local degrees of freedom. Actually, these all belong
to the gauge degrees of freedom.

Hence, it is crucial to find  a relevant gravity theory which has
a physically massive mode propagating on the Minkowski spacetime.
One model is given by the topologically new massive gravity
(TNMG)~\cite{Andringa:2009yc,Dalmazi:2009pm}. Very recently, we
have shown that the linearized TNMG provides a single spin-2 mode
with mass $m^2/\mu$ in the Minkowski spacetime~\cite{Kim:2013yu}.
On the other hand, it is well known that  the cosmological
generalized massive gravity (cGMG) has two different massive
modes~\cite{Bergshoeff:2009hq,Bergshoeff:2009aq}. Furthermore, the
cGMG provides the standard rank-2 LCFT for $m_1\ell =1$, while it
yields the exotic rank-2 LCFT for $m_1\ell=m_2\ell
\not=1$~\cite{Grumiller:2010tj}.

In this work, we explore the critical GMG where two masses
degenerate in the Minkowski spacetime. We obtain the massive
logarithmic wave solution and  discuss the logarithmic extension of
the GCA which may arise from the exotic rank-2 LCFT on the boundary
of the AdS$_3$ spacetime. Furthermore, we show that in the
flat-spacetime limit of $\ell \to \infty$, the cGMG for $m_1\ell =1$
reduces to a logarithmic GCA in the semi-infinite spacetime with a
boundary condition at $\phi=0$, while the critical GMG yields null
in the limit of $\ell \to \infty$. In other words, the standard
rank-2 LCFT survives in the flat-spacetime limit, whereas the exotic
rank-2 LCFT disappears in the flat-spacetime limit.

\section{GMG and its  critical  theory}
The GMG action is given
by~\cite{Bergshoeff:2009hq,Bergshoeff:2009aq}
 \be \label{gmg}
 I_{\rm GMG}=\frac{1}{\kappa^2}\int d^3x\sqrt{-g} (\sigma R)+I_{\rm CSG}+ I_{\rm K},
 \ee
where
 \ba
 && I_{\rm CSG} = \frac{1}{2\kappa^2\mu}\int\!\! d^3x\sqrt{-g} \epsilon^{\lambda\mu\nu}
             \Gamma^\rho_{\lambda\sigma}
             \!\!\left(\partial_\mu\Gamma^\sigma_{\rho\nu}+
                   \frac{2}{3}\Gamma^\sigma_{\mu\tau}\Gamma^\tau_{\nu\rho}\right),
                   \nonumber \\
 && I_{\rm K} = \frac{1}{\kappa^2m^2}\int\!\! d^3x\sqrt{-g}
             \left(R^{\mu\nu}R_{\mu\nu}-\frac{3}{8}R^2\right)\equiv  \frac{1}{\kappa^2m^2}\int\!\! d^3x\sqrt{-g} K.
 \ea
Here $\kappa^2=16\pi G$ is the three-dimensional gravitational
coupling, $\mu$ the Chern-Simons coupling, and $m^2$ the mass
parameter for a higher curvature combination $K$. $\sigma$ is
chosen to be 1 for our purpose. The Einstein  equation of the GMG
action is given by
 \be\label{eom}
 \sigma G_{\mu\nu}+\frac{1}{\mu}C_{\mu\nu}+\frac{1}{2m^2}K_{\mu\nu}=0,
 \ee
where the Einstein tensor $G_{\mu\nu}$ is
 \be
 G_{\mu\nu}=R_{\mu\nu}-\frac{1}{2} g_{\mu\nu} R,
 \ee
the Cotton tensor $C_{\mu\nu}$ takes the form \be
C_{\mu\nu}=\epsilon_\mu^{~\alpha\beta}\nabla_\alpha
            \left(R_{\beta\nu}-\frac{1}{4}g_{\beta\nu}R\right),
 \ee
and the tensor $K_{\mu\nu}$ is given by
 \ba
 K_{\mu\nu}&=&2\nabla^2R_{\mu\nu}-\frac{1}{2}\nabla_\mu \nabla_\nu R-\frac{1}{2}\nabla^2Rg_{\mu\nu}
          +4R_{\mu\rho\nu\sigma}R^{\rho\sigma} \nonumber\\
        &-&\frac{3}{2} R R_{\mu\nu}-g_{\mu\nu}R_{\rho\sigma}R^{\rho\sigma}
         +\frac{3}{8}{R}^2g_{\mu\nu}.
 \ea

As a solution to the Einstein equation  (\ref{eom}), the Minkowski
spacetime is chosen to be
 \be \label{EF}
 ds^2_{\rm EF}=\bar{g}_{\mu\nu}dx^\mu dx^\nu= -du^2-2drdu+r^2d\theta^2,
 \ee
where $u=t-r$ is a retarded time expressed in terms of the outgoing
Eddington-Finkelstein (EF) coordinates.  In this spacetime,
non-vanishing Christoffel symbols are given by
 \be
 \Gamma^u_{\theta\theta}=r,~~\Gamma^r_{\theta\theta}=-r,
 ~~\Gamma^\theta_{r\theta}=\Gamma^\theta_{\theta r}=\frac{1}{r},
 \ee
while all the components of curvature  tensor $R_{\mu\nu\rho\sigma}$
are zero.

Considering the perturbation $h_{\mu\nu}$ around the EF background
$\bar{g}_{\mu\nu}$
 \be
 g_{\mu\nu}=\bar{g}_{\mu\nu}+h_{\mu\nu},
 \ee
the linearized equation of Eq.~(\ref{eom}) takes the form
 \be
\sigma \delta G_{\mu\nu}(h)+ \frac{1}{\mu}\delta
C_{\mu\nu}(h)+\frac{1}{2m^2}\delta K_{\mu\nu}(h)=0,
 \ee
where linearized tensors are given in~\cite{Liu:2009pha} with
$\Lambda=0$. It is necessary to choose the transverse and
traceless gauge conditions to select two  massive modes
propagating on the EF background as
 \be \label{ttcon}
 \bar{\nabla}^\mu h_{\mu\nu}=0,~~~h^\mu_{~\mu}=0.
 \ee
Then, we have the linearized fourth-order  equation of motion as
 \be \label{lineq}
\bar{\nabla}^2
 \Bigg[\bar{\nabla}^2h_{\mu\nu}+\frac{m^2}{\mu}\epsilon^{~\alpha\beta}_\mu\bar{\nabla}_\alpha h_{\beta\nu}
+\sigma m^2h_{\mu\nu}\Bigg]=0.
 \ee
Introducing  three mutually commuting operators as
 \be
 D^\beta_\mu=\epsilon_\mu^{~\alpha\beta}\bar{\nabla}_\alpha,
~~~(D^{m_i})^\beta_\mu=\delta^\beta_\mu+\frac{1}{m_i}\epsilon_\mu^{~\alpha\beta}\bar{\nabla}_\alpha,~~i=1,2,
 \ee
the fourth order equation (\ref{lineq})  can be expressed
 \be\label{lin-eom}
\left(D^2D^{m_1}D^{m_2} h\right)_{\mu\nu}=0,
 \ee
where two masses are defined by \be \label{masses}
 m_1=\frac{m^2}{2\mu}+\sqrt{\frac{m^4}{4\mu^2}-\sigma m^2},~~
 m_2=\frac{m^2}{2\mu}-\sqrt{\frac{m^4}{4\mu^2}-\sigma
 m^2}.
 \ee
We note that $D$ is the massless operator, while $D^{m_i}$
correspond to two massive operators.  Two masses are positive for
$m^2> 4\mu^2$ with $\sigma=1$. For $m^2=4\mu^2$, two masses
degenerate as \be m_1=m_2=2\mu \equiv m_0, \ee  which leads to a
critical GMG in the flat spacetime.

In order to obtain the wave solution to the critical GMG, we first
have to solve the first-order massive equation
 \be\label{meq}
 (D^{m_i}h)_{\mu\nu}
 =h_{\mu\nu}+\frac{1}{m_i}\epsilon_\mu^{~\alpha\beta}\bar{\nabla}_\alpha h_{\beta\nu}=0
 \ee
under the traceless and transverse gauge conditions (\ref{ttcon}).
Since the massive solution was already found for the case of
single degree of freedom~\cite{Kim:2013yu}, we could write down
the solution for the case of double degrees of freedom as follows
 \be \label{tms}
 h^{m_i}_{\mu\nu}(u,r,\theta)=e^{-im_i(u+r)}e^{-2i\theta}
  \left(
  \begin{array}{ccc}
   0 & 0 & 0 \\
    0 & 1 & -ir \\
    0 & -ir & -r^2 \\
  \end{array}
 \right)
 \equiv e^{-im_i(u+r)}e^{-4i\theta}(m_{1}\bigotimes m_{1})_{\mu\nu}.
 \ee
Then, at the critical point, the Einstein equation has reduced to
 \be\label{ceq}
 \Big[(D^{2\mu})^2h^{\rm c}\Big]_{\mu\nu}=0.
 \ee
Its solution is found to  be  \be \label{cmg}h^{\rm
c}_{\mu\nu}=\partial_{m_i} h^{m_i}_{\mu\nu}|_{m_i\to 2\mu}=-i(u+r)
h^{2\mu}_{\mu\nu}, \ee which is our main result. Here we observe
that there is no log-term in the critical tensor wave solution
$h^{\rm c}_{\mu\nu}$ in the flat spacetime, in comparison to the
logarithmic solution $\psi^{\rm log}_{\mu\nu}=-[i \tau +\ln \cosh
\rho]\psi^{\rm L}_{\mu\nu}$ on the AdS$_3$ spacetime.

\section{Re-deriving the critical wave solution}

It is important to justify the critical wave solution (\ref{cmg})
because it was newly derived in the flat spacetime.  In order to
re-derive  (\ref{cmg}), we first consider the AdS/CFT
correspondence on the AdS$_3$ and its boundary. The global AdS$_3$
spacetime could be described by the line element
 \be\label{global}
 ds^2_{\rm AdS_3}=\ell^2 \Big(-\cosh^2\rho d\tau^2+d\rho^2+\sinh^2\rho d\phi^2\Big),
 \ee
where all coordinates are defined to be dimensionless.

Two central charges of the cGMG on the boundary are given
by~\cite{Liu:2009pha,Bergshoeff:2012ev,Kim:2012rz},
 \be
 c_L=\frac{3\ell}{2G}\Big(\sigma+\frac{1}{2m^2\ell^2}-\frac{1}{\mu\ell}\Big),
 ~c_R=\frac{3\ell}{2G}\Big(\sigma+\frac{1}{2m^2\ell^2}+\frac{1}{\mu\ell}\Big).
 \ee
To obtain the GMG in the flat limit of $\ell \to \infty$, the
corresponding BMS central charges are defined to be
 \be c_1=\lim_{\ell \to \infty}(c_R-c_L)=\frac{3}{G\mu},
 ~~c_2=\lim_{\ell \to\infty}\frac{c_R+c_L}{\ell}=\frac{3\sigma}{G},
 \ee
which show the absence of the mass parameter $m^2$ in the flat
spacetime limit. We introduce the scaling limits of $\mu \to 0$
and $G \to \infty$ while keeping fixed $\mu G$.  Considering a
relation $G\mu=1/8k$, its dual CFT is given by the 2D GCA  with
central charges~\cite{Bagchi:2012yk}
 \be
 c_1=24k,~~c_2=0.
 \ee
Considering the highest weight condition of the cGMG on the
AdS$_3$ spacetime: ${\cal L}_0|\psi_{\mu\nu}>=h|\psi_{\mu\nu}>$
and $\bar{\cal L}_0|\psi_{\mu\nu}>=\bar{h}|\psi_{\mu\nu}>$,   the
connection between the GCA and the Virasoro algebras  is given by
 \be
 L_n={\cal L}_n-\bar{\cal L}_{-n},
 ~~M_n=\frac{\bar{\cal L}_n+{\cal L}_{-n}}{\ell}.
 \ee
Here $h$ and $\bar{h}$ are given for the cGMG
by~\cite{Grumiller:2010tj}
 \be\label{hh}
 (h_{1/2},\bar{h}_{1/2})=\Big(\frac{3+\ell \tilde{m}_{1/2}}{2}, \frac{-1+\ell \tilde{m}_{1/2}}{2}\Big)
 \ee
where
 \be\label{m-gmg}
 \tilde{m}_{1/2}=\frac{m^2}{2\mu}\pm \sqrt{\frac{1}{2\ell^2}+\frac{m^4}{4\mu^2}-\sigma
 m^2}.
 \ee
Since the rigidity $\xi$ and scaling dimension $\Delta$ are
eigenvalues as
 \be
 L_0|\triangle,\xi>=\xi|\triangle,\xi>,~~
 M_0|\triangle,\xi>=\triangle |\triangle,\xi>,
 \ee
they  are given  by
 \be \xi=\lim_{\ell \to \infty}(h-\bar{h}),
 ~~\triangle=\lim_{\ell \to \infty}\frac{h+\bar{h}}{\ell}.
 \label{xiad}\ee
We obtain
 \be\label{fvalue}
 \xi=2,~~ \triangle=m_i,
 \ee
where $m_i$ is already given by Eq.~(\ref{masses}).  The
eigenvalue $\xi=2$ arises because it represents spin-2
excitations.

Now, it is very interesting to know what form of the
GMG~\cite{Liu:2009pha} provides Eq.~(\ref{cmg}) in the
flat-spacetime limit. For the condition of
 \be
 \frac{1}{2\ell^2}+\frac{m^4}{4\mu^2}-\sigma m^2=0,
 \ee
one has the degenerate mass from Eq.~(\ref{m-gmg})
 \be
 \tilde{m}_0=\frac{m^2}{2\mu},
 \ee
which provides the exotic rank-2 LCFT on the boundary of the
AdS$_3$ spacetime~\cite{Grumiller:2010tj}. In this case, the
critical GMG wave solution
 in the light-cone coordinate is
described by
 \be
 \psi^{\rm elog}_{\mu\nu} (\rho, \tau^+, \tau^-)
  =\partial_{\tilde{m}_i}\psi^{\rm \tilde{m}_i}_{\mu\nu}|_{\tilde{m}_i \to \tilde{m}_0}
  = y(\tau,\rho)\psi^{\rm \tilde{m}_0}_{\mu\nu} (\rho, \tau^+, \tau^-)
 \ee
where the logarithmic function $y$ is
 \be
 y(\tau,\rho)=-[i\tau+\ln \cosh \rho]\ell
 \ee
and the massive wave function is
 \be
 \psi^{\rm \tilde{m}_0}_{\mu\nu} (\rho, \tau^+, \tau^-) = f(\rho, \tau^+, \tau^-)
 \left(
  \begin{array}{ccc}
   1 & 0 & \frac{2i}{\sinh(2\rho)} \\
   0 & 0 & 0 \\
    \frac{2i}{\sinh(2\rho)} & 0 & -\frac{4}{\sinh^2(2\rho)}  \\
  \end{array}
 \right)
 \ee
with $\tau^\pm=\tau \pm\phi$. Here
 \be
 f(\rho,\tau^+,\tau^-)=e^{-ih(\tilde{m}_0)\tau^+-i\bar{h}(\tilde{m}_0)\tau^-}
               (\cosh\rho)^{-[h(\tilde{m}_0)+\bar{h}(\tilde{m}_0)]}\sinh^2\rho.
 \ee
We express the global coordinates (\ref{global}) in terms of the
EF coordinates (\ref{EF}) in Ref.~\cite{Bagchi:2012yk}
 \be
 u=\ell(\tau-\rho),~~r=\ell\rho,~~\theta=\phi.
 \ee
Taking the limit of $\ell\rightarrow\infty$ while keeping $u$ and
$r$ finite and  making use of Eq.~(\ref{xiad}), we arrive at
 \be
 \psi^{\rm elog}_{\mu\nu} (u,r,\theta)=-i(u+r) \psi^{\rm \tilde{m}_0}_{\mu\nu}(u,r,\theta),
 \ee
where
 \be \label{atof}
 \psi^{\rm \tilde{m}_0}_{\mu\nu}(u,r,\theta)\simeq e^{-i2 \mu (u+r) -2i\theta}
 \left(
  \begin{array}{ccc}
   0 & 0 & 0 \\
   0 & 1 & -ir \\
   0 & -ir & -r^2  \\
  \end{array}
 \right).
 \ee
It shows clearly that the massive wave solution (\ref{cmg})
represents a critical massive graviton mode propagating in the
Minkowski spacetime background.

On the other hand, we note here that the GCA generator  $M_0$
acquires a rank-2 Jordan cell as
 \be \label{mzero}
 M_0 \left(
  \begin{array}{c}
   \psi^{\rm elog}_{\mu\nu}  \\
   \psi^{\rm \tilde{m}_0}_{\mu\nu}  \\
  \end{array}
 \right)
  =
   \left(
  \begin{array}{cc}
   2\mu & 1 \\
   0 & 2\mu \\
  \end{array}
 \right)
 \left(
  \begin{array}{c}
   \psi^{\rm elog}_{\mu\nu}  \\
   \psi^{\rm \tilde{m}_0}_{\mu\nu}  \\
  \end{array}
 \right),
 \ee
 where $2\mu$ is identified  with the scaling dimension $\Delta$.
\section{Is Logarithmic GCA possible?}

As was shown previously, the BMS/GCA correspondence for the GMG
provides the GCA$_2$ with the central charges $c_1=24k$ and
$c_2=0$. What is the corresponding GCA to the critical GMG? It
seems to be existed because of the presence of the rank-2 Jordan
cell (\ref{mzero}). In order to answer this question, we review
the exotic rank-2 LCFT on the AdS$_3$ spacetime. The exotic rank-2
LCFT is composed of two operators $\{{\cal O}(z), {\cal O}^{\rm
log}(z)\}$. The two-point functions of these operators take the
forms~\cite{Grumiller:2010tj}
 \ba
 <{\cal O}^1(z,\bar{z}){\cal O}^1(0,0)>&=&0, \label{lcft1}\\
 <{\cal O}^{\rm log}(z,\bar{z}){\cal O}^1(0,0)>
  &=&\frac{B}{2z^{2h(\tilde{m}_0)}\bar{z}^{2\bar{h}(\tilde{m}_0)}},\label{lcft2}\\
 <{\cal O}^{\rm log}(z,\bar{z}){\cal O}^{\rm log}(0,0)>
  &=&-\frac{B\ln(m_L
  |z|^2)}{2z^{2h(\tilde{m}_0)}\bar{z}^{2\bar{h}(\tilde{m}_0)}},\label{lcft3}
 \ea
where $B$ and $m_L$ are non-zero parameters. We note that $<{\cal
O}^1(z,\bar{z}){\cal O}^1(0,0)>$ is null $(c_{\rm \tilde{m}_0}=0)$
for a LCFT. Here we use  the relations
 \be
(h,\bar{h})=\left(\frac{3}{2}+\frac{\ell \tilde{m}_0}{2},
  ~-\frac{1}{2}+\frac{\ell \tilde{m}_0}{2}\right).
 \ee
In order to match  the flat-spacetime limit, we introduce two
coordinates
 \be \label{z-coo}
 z=t+\frac{1}{\ell}\phi,
 ~~ \bar{z}=t-\frac{1}{\ell}\phi
 \ee
with $t=u+r$. In the flat-spacetime limit of $\ell \to \infty$, we
recall that two expression  of $h-\bar{h}=2$ and $\lim_{\ell \to
\infty} (h+\bar{h})/\ell=\Delta $ in Eq.~(\ref{xiad}) are finite.
In the non-relativistic limit of $(z,\bar{z} \to \tau,~\phi \to
0)$ induced by the flat-spacetime limit of $\ell \to \infty$
~\cite{Setare:2012ks}, the relations (\ref{lcft2}) and
(\ref{lcft3}) reduce to
 \ba\label{two1}
 <{\cal O}^{\rm log}(z,\bar{z}){\cal O}^1(0,0)>
  &\sim&\frac{1}{2t^{\infty}}\sim 0,\\
  \label{two-log}
 <{\cal O}^{\rm log}(z,\bar{z}){\cal O}^{\rm log}(0,0)>
  &\sim&-\frac{\ln(t)}{t^{\infty}} \sim 0
 \ea
in the semi-infinite spacetime with a boundary condition at
$\phi=0$.  For $t\not=0$, these are null.  This implies that the
flat spacetime limit of the exotic rank-2 LCFT is not properly
defined.

For a massive scalar
$\Phi$~\cite{Setare:2012ks,Alishahiha:2009np}, its corresponding
two-point function of GCA$_2$ is given by \be <{\cal O}(t){\cal
O}(0)> \sim \frac{1}{t^{2\Delta}}, \ee with
$\Delta=\sqrt{m^2+1}+1/2$.

Finally, we investigate the flat-spacetime limit of the standard
rank-2 LCFT which corresponds  to the critical case of
$\tilde{m}_1\ell=1$. The non-vanishing two-point correlators are
given by
 \ba
 <{\cal O}^{\rm log}(z,\bar{z}){\cal O}^L(0,0)> &=&\frac{b_L}{2z^{4}},\label{slcft1}\\
 <{\cal O}^{\rm log}(z,\bar{z}){\cal O}^{\rm log}(0,0)> &=&-\frac{b_L\ln(m_L |z|^2)}{z^{4}}, \label{slcft2}
 \ea
 where $b_L$ and $m_L$ are non-zero parameters and the zero left central charge,
 $c_L=0$. Also, one has $h=2$ and $ \bar{h}=0$.
Choosing Eq.~(\ref{z-coo}) and taking the non-relativistic  limit,
the relations (\ref{slcft1}) and (\ref{slcft2}) reduce to
 \ba\label{two11}
 <{\cal O}^{\rm log}(z,\bar{z}){\cal O}^L(0,0)>
  &\sim&\frac{1}{2t^{4}},\\
  \label{two-log1}
 <{\cal O}^{\rm log}(z,\bar{z}){\cal O}^{\rm log}(0,0)>
  &\sim&-\frac{\ln(t)}{t^{4}}
 \ea
in the semi-infinite spacetime with a boundary condition at
$\phi=0$. The relations (\ref{two11}) and (\ref{two-log1}) have
been dubbed a logarithmic GCA$_2$ as the non-relativistic limit of
rank-2 LCFT~\cite{Hosseiny:2010sj,Hosseiny:2011ct,Hyun:2012fd}.
However, there is a difference between logarithmic Schr\"odinger
invariance and logarithmic Galilean conformal
invariance~\cite{Henkel:2012hd}.
 In this case, two-point
correlation functions are represented by a matrix representation
 \be
 <{\cal O}^i{\cal O}^j> \sim
 \left(
  \begin{array}{cc}
   0 & {\rm CFT} \\
   {\rm CFT} & {\rm Log}   \\
  \end{array}
 \right),
 \ee
where $i,j=L, {\rm log}$,  CFT denotes the CFT two-point function
(\ref{two11}), and  Log represents (\ref{two-log1}). In this case,
the critical wave solution to $D^2h^{\rm log}_{\mu\nu}=0$ is given
by $ h^{\rm log}_{\mu\nu}=-i t h^\xi_{\mu\nu}$.

\section{Discussions}

We have obtained the massive logarithmic wave solution from the
critical GMG in three dimensional flat spacetime. In contrast to
the logarithmic wave solution on the AdS$_3$ spacetime, it does
not contain a logarithmic term in the flat spacetime.

At the off-critical point of GMG, the dual CFT is given by the
GCA$_2$ with central charges $c_1=24k$ and $c_2=0$, where the mass
parameter $m$ disappears in the flat-spacetime limit.

On the other hand, at the critical point of GMG, we have discussed
the logarithmic extension of the GCA  which may arise from the
exotic rank-2 LCFT on the boundary of the AdS$_3$ spacetime. The
``exotic" rank-2 LCFT implies that it appears when two masses
degenerate on the boundary of the AdS$_3$ spacetime.
 It turned out that the logarithmic GCA$_2$ is not realized from the
exotic rank-2 LCFT. However, the logarithmic GCA$_2$ could be
realized from the standard rank-2 LCFT when taking the flat
spacetime limit.

\begin{acknowledgments}
 This
work was supported by the National Research Foundation of Korea
(NRF) grant funded by the Korea government (MEST) through the
Center for Quantum Spacetime (CQUeST) of Sogang University with
grant number 2005-0049409. Y. S. Myung was also supported by the
National Research Foundation of Korea (NRF) grant funded by the
Korea government (MEST) (No.2012-040499). Y.-J. Park was also
supported by World Class University program funded by the Ministry
of Education, Science and Technology through the National Research
Foundation of Korea(No. R31-20002).
\end{acknowledgments}


\begin{thebibliography}{99}



\bibitem{Bagchi:2012yk}
  A.~Bagchi, S.~Detournay and D.~Grumiller,
  Phys.\ Rev.\ Lett.\  {\bf 109}, 151301 (2012)  [arXiv:1208.1658 [hep-th]].

\bibitem{Bagchi:2010eg}
  A.~Bagchi,
  Phys.\ Rev.\ Lett.\  {\bf 105}, 171601 (2010) [arXiv:1006.3354 [hep-th]].



\bibitem{Barnich:2006av}
  G.~Barnich and G.~Compere,
   Class.\ Quant.\ Grav.\  {\bf 24}, F15 (2007)  [gr-qc/0610130].

\bibitem{Bagchi:2009pe}
  A.~Bagchi, R.~Gopakumar, I.~Mandal and A.~Miwa,
  JHEP {\bf 1008}, 004 (2010)  [arXiv:0912.1090 [hep-th]].



\bibitem{Andringa:2009yc}
  R.~Andringa, E.~A.~Bergshoeff, M.~de Roo, O.~Hohm, E.~Sezgin and P.~K.~Townsend,
  Class.\ Quant.\ Grav.\  {\bf 27}, 025010 (2010)  [arXiv:0907.4658 [hep-th]].

\bibitem{Dalmazi:2009pm}
  D.~Dalmazi and E.~L.~Mendonca,
   JHEP {\bf 0909}, 011 (2009)  [arXiv:0907.5009 [hep-th]].

\bibitem{Kim:2013yu}
  Y.~-W.~Kim, Y.~S.~Myung and Y.~-J.~Park,
    arXiv:1301.0969 [hep-th].

\bibitem{Bergshoeff:2009hq}
  E.~A.~Bergshoeff, O.~Hohm and P.~K.~Townsend,
  Phys.\ Rev.\ Lett.\  {\bf 102}, 201301 (2009)  [arXiv:0901.1766 [hep-th]].

\bibitem{Bergshoeff:2009aq}
  E.~A.~Bergshoeff, O.~Hohm and P.~K.~Townsend,
   Phys.\ Rev.\ D {\bf 79}, 124042 (2009)  [arXiv:0905.1259 [hep-th]].



\bibitem{Grumiller:2010tj}
  D.~Grumiller, N.~Johansson and T.~Zojer,
   JHEP {\bf 1101}, 090 (2011)  [arXiv:1010.4449 [hep-th]].


\bibitem{Liu:2009pha}
  Y.~Liu and Y.~-W.~Sun,
   Phys.\ Rev.\ D {\bf 79}, 126001 (2009)  [arXiv:0904.0403 [hep-th]].

\bibitem{Bergshoeff:2012ev}
  E.~A.~Bergshoeff, S.~de Haan, W.~Merbis, J.~Rosseel and T.~Zojer,
  Phys.\ Rev.\ D {\bf 86}, 064037 (2012)  [arXiv:1206.3089 [hep-th]].


\bibitem{Kim:2012rz}
  Y.~-W.~Kim, Y.~S.~Myung and Y.~-J.~Park,
   Phys.\ Rev.\ D {\bf 86}, 064017 (2012)  [arXiv:1207.3149 [hep-th]].



\bibitem{Setare:2012ks}
  M.~R.~Setare and V.~Kamali,
   Eur.\ Phys.\ J.\ C {\bf 72}, 2115 (2012)  [arXiv:1202.4917 [hep-th]].


\bibitem{Alishahiha:2009np}
  M.~Alishahiha, A.~Davody and A.~Vahedi,
  JHEP {\bf 0908}, 022 (2009)  [arXiv:0903.3953 [hep-th]].


\bibitem{Hosseiny:2010sj}
  A.~Hosseiny and S.~Rouhani,
   J.\ Math.\ Phys.\  {\bf 51}, 102303 (2010)  [arXiv:1001.1036 [hep-th]].

\bibitem{Hosseiny:2011ct}
  A.~Hosseiny and A.~Naseh,
   J.\ Math.\ Phys.\  {\bf 52}, 092501 (2011)  [arXiv:1101.2126 [hep-th]].

\bibitem{Hyun:2012fd}
  S.~Hyun, J.~Jeong and B.~S.~Kim,
  arXiv:1209.2417 [hep-th].


\bibitem{Henkel:2012hd}
  M.~Henkel,
    arXiv:1205.5901 [math-ph].

 \end{thebibliography}
 \end{document}